\documentclass[
preprint,
superscriptaddress,
amsmath,amssymb,
showkeys,
aps,
prb,
]{revtex4-2}

\usepackage{threeparttable}
\usepackage{setspace}
\usepackage{makecell}
\usepackage{amsmath}
\usepackage{graphicx}
\usepackage{dcolumn}
\usepackage{bm}
\usepackage{hyperref}
\usepackage[mathlines]{lineno}
\usepackage{textcomp}
\usepackage{gensymb}
\usepackage{subfigure}
\usepackage{url}
\usepackage{xcolor}
\usepackage{hhline}
\usepackage[version=3]{mhchem}
\usepackage{multirow}
\usepackage[american]{babel}
\usepackage{hhline}
\usepackage{textgreek}
\usepackage{miller}

\definecolor{red}{rgb}{0.6,0.1,0.1}\def\red{\color{red}}
\definecolor{blue}{rgb}{0.1,0.1,0.6}
\definecolor{green}{rgb}{0.1,0.6,0.1}
\definecolor{yellow}{rgb}{0.8,0.8,0.6}

\definecolor{JL}{rgb}{0.0,0.0,0.0}\def\JL{\color{JL}}

\definecolor{HH}{rgb}{0.0,0.0,0.0}\def\HH{\color{HH}}

\definecolor{FI}{rgb}{0.0,0.0,0.0}

\newcommand{\FD}[1]{\textcolor{black}{#1}}

\begin{document}


\title{Threshold displacement energy map of Frenkel pair generation in \texorpdfstring{\textit{\textbeta}-\ce{Ga2O3}}{} from machine-learning-driven molecular dynamics simulations}

\author{Huan He} 
\affiliation{School of Nuclear Science and Technology, Xi'an Jiaotong University, Xi'an, Shaanxi 710049, China}
\affiliation{Department of Physics, University of Helsinki, P.O. Box 43, FI-00014, Finland}

\author{Junlei Zhao} \email{zhaojl@sustech.edu.cn}
\affiliation{Department of Electrical and Electronic Engineering, Southern University of Science and Technology, Shenzhen 518055, China}

\author{Jesper Byggm{\"a}star} 
\affiliation{Department of Physics, University of Helsinki, P.O. Box 43, FI-00014, Finland}

\author{Ru He} 
\affiliation{Department of Physics, University of Helsinki, P.O. Box 43, FI-00014, Finland}

\author{Kai Nordlund} 
\affiliation{Department of Physics, University of Helsinki, P.O. Box 43, FI-00014, Finland}
\affiliation{Helsinki Institute of Physics, University of Helsinki, P.O. Box 43, FI-00014, Finland}

\author{Chaohui He} 
\affiliation{School of Nuclear Science and Technology, Xi'an Jiaotong University, Xi'an, Shaanxi 710049, China}

\author{Flyura Djurabekova} \email{flyura.djurabekova@helsinki.fi}
\affiliation{Department of Physics, University of Helsinki, P.O. Box 43, FI-00014, Finland}
\affiliation{Helsinki Institute of Physics, University of Helsinki, P.O. Box 43, FI-00014, Finland}


\begin{abstract}

{\JL $\beta$-gallium oxide} ($\beta$-\ce{Ga2O3}) shows great promise 
for electronics applications, particularly, in future space operating devices exposed to harsh radiation environments for extended times. 
 This study focuses on crucial, yet not fully explored, aspects of radiation damage in this material, such as threshold displacement energies 
and formation of various radiation-induced 
Frenkel pairs. Analyzing 
over 5,000 molecular dynamics simulations based on our machine-learning potentials, we 
conclude that the threshold displacement energies for two Ga sites, the tetrahedral (22.9 eV) and octahedral (20 eV) ones, differ stronger than the same values for three different O sites, which range only between 17 eV and 17.4 eV.
Mapping of threshold displacement energies unveils significant differences in displacements for all five {\JL atomic sites}. Our newly developed defect identification methodology successfully classified multiple Frenkel pair types in $\beta$-\ce{Ga2O3}, with over ten different Ga and two primary O ones with 
a predominant O split interstitial at the O1 site. Finally, the calculated recombination energy barriers suggest that O Frenkel pairs are more likely to recombine {\JL upon annealing} than Ga. These insights are pivotal for understanding the radiation damage and defect formation in \ce{Ga2O3}, providing the basis for design of \ce{Ga2O3}-based electronics with high radiation resistance.

\end{abstract}
\maketitle
\newpage
\section{Introduction}   

$\beta$-gallium oxide ($\beta$-\ce{Ga2O3}) is the most stable polymorph among the {\JL five experimentally known polymorphic phases} of \ce{Ga2O3}~\cite{review-wang2022recent,multipolymorphs}.
An ultra-wide bandgap (4.8-4.9 eV), a remarkably high breakdown electric field (8 MV/cm), and a remarkably large Baliga's figure of merit ($\mathrm{BFOM}=\epsilon \mu E_\mathrm{crit.}^{3}=3444$) make $\beta$-\ce{Ga2O3} a highly promising candidate material for high-power electronics~\cite{mastro2017perspective, tsao2018ultrawide, qiaoStateofartReviewGallium2022} and solar-blind ultraviolet detectors operating in the wavelength range of 220-280 nm~\cite{detector, qian-detector}.


{\JL Owing} to its {\JL strong ionic bonds} and high radiation resistance, \ce{Ga2O3}-based devices \FD{can also be considered for integration, \textit{e.g.}, in future spacecrafts to operate in harsh environments with limited radiation protection}~\cite{kim2019radiation}. 
{\JL On the other hand,} during the fabrication process, ion implantation is {\JL commonly} employed to modify the (opto-)electronic properties of \ce{Ga2O3}, especially {\JL concerning} the current challenge of {\JL effective $p$-type doping}. 
{\JL Consequently, extensive studies have focused on exploring the radiation damage of {\JL $\beta$-\ce{Ga2O3}} in recent years, and have shown that \ce{Ga2O3} exhibits considerable radiation hardness compared to other semiconductor materials~\cite{peartonReviewRadiationDamage2021,yang20171,wong2018radiation}.}
\FD{The recently reported phase transition from $\beta$- to $\gamma$-\ce{Ga2O3} upon heavy ion irradiation~\cite{azarovUniversalRadiationTolerant2023}, which maintains its crystallinity up to exceptionally high radiation doses also indicates the unprecedentedly high radiation resistance of this material, compared to any other currently used in the semiconductor industry.}


Despite strong interest to this material, the atomic-level mechanisms behind this behavior remain poorly understood. 
Typically, when a high-energy particle impacts on a solid-state material, it transfers the energy to the target atoms upon collisions. 
The atoms displaced in collision cascades may result in creation of \FD{Frenkel pairs (FPs), \textit{i.e.}, interstitial-vacancy pairs characterizing single displaced atoms. }
The value of energy threshold that is required to generate \FD{a single FP}, also known as the {\JL threshold displacement energy (TDE)}, $E_\mathrm{d}$, 
can be used to estimate the level of radiation damage generated by a given ion with a given energy~\cite{nordlund2006molecular} using NRT~\cite{NRT-model-norgett1975proposed}, Kinchin-Pease~\cite{kp-model-kinchin1955displacement} or arc-dpa~\cite{nordlund2018primary} models.  While \textit{in-situ} experimental techniques have made significant progress in observing radiation damage processes, capturing displacement events remains challenging due to their ultra-short timescales (typically of the order of picoseconds).
In contrast, theoretical calculations often offer a deeper understanding of the atomic-scale mechanisms involved. 
Such simulations have been successfully employed to accurately calculate $E_\mathrm{d}$ and simulate the displacement damage for various materials, such as SiC~\cite{jiangComparativeStudyLow2016}, GaN~\cite{nordMolecularDynamicsStudy2003, xiao2009threshold, hePrimaryDamage102020}, and AlN~\cite{xiInitioMolecularDynamics2018}. 

Two prevailing methods, namely \textit{ab initio} molecular dynamics (AIMD)~\cite{jiangComparativeStudyLow2016,xiInitioMolecularDynamics2018, xiao2009threshold} and classical molecular dynamics (MD)~\cite{byggmastar2018effects, Jesper-byggmastar2019machine, nordMolecularDynamicsStudy2003,hePrimaryDamage102020}, are commonly utilized to calculate $E_\mathrm{d}$ values. 
AIMD simulations based on density functional theory (DFT) enable the observation of the entire dynamic process but suffer from limitations to small simulation sizes (typically a few hundreds of atoms) and short time scales due to computational speed restrictions. 
For instance, a recent study of $E_\mathrm{d}$ in $\beta$-\ce{Ga2O3} performed using AIMD simulations~\cite{AIMD-2023TDEGa2O3} reported fairly low values $28\pm1$ eV for \ce{Ga} and $14\pm1$ eV for \ce{O}.  
\FD{We also note here that the TDE value is strongly crystal direction-dependent, hence, sufficient sampling of random displacement directions is required for more accurate value of TDE.}
Therefore, classical MD simulations are more efficient as they can simulate a large system and {\JL a vast number of displacement} directions, providing statistically significant results. 
However, the accuracy of classical MD entirely relies on {\JL the accuracy of} the interatomic {\JL potential (IAP)} that is used to describe interatomic interactions. 
Describing the atomic forces in $\beta$-\ce{Ga2O3} is highly challenging for the \FD{classical analytical} IAPs due to the five nonequivalent atomic types and low-symmetry nature of the system. 
Thanks to the development of machine-learning algorithms, more suitable machine-learning IAPs (ML-IAPs) became available~\cite{Jesper-byggmastar2019machine, mlIP-graphene, zhao2023complex} for simulations of 
more complicated material systems, which enabled the simulations of multi-million atomic systems with quantum-mechanical level of accuracy.

In this study, we report 
the results of comprehensive investigation of TDE values 
in $\beta$-\ce{Ga2O3} using ML-MD method. 
More than 5,000 directions are simulated in total to obtain {\JL statistically significant} values. 
Moreover, we develop a method to detect and classify the specific types of FPs accurately. 
The properties and the recombination behavior of FPs are explored using various analytical methods. 
Furthermore, we compare and discuss our results with previous theoretical and experimental studies. 
Our findings contribute to a more profound understanding of the radiation effects of this emerging semiconductor material at the atomic scale.

\newpage

\section{Methodology} 

\subsection{Threshold displacement energies simulation} \label{sec:tde_method}

All TDE calculations are performed by the Large-scale Atomic/Molecular Massively Parallel Simulator (LAMMPS) code~\cite{LAMMPS} and QUIP package~\cite{quip, Bartok2010-pwGAPcode}. 
The interactions between atoms in \ce{Ga2O3} are described by our two versions of the newly developed ML-IAPs, namely soapGAP and tabGAP~\cite{zhao2023complex}. 
The database used to train the two ML-IAPs consists of over 1,600 structures containing more than 100,000 local atomic environments and provides a {carefully refitted} repulsive potential for the short-range atomic interactions to simulate the high-energy recoil events and defect evolution. 
The main difference between the soapGAP and tabGAP is their accuracy and computational speed. 
\FD{The accuracy of the simulation results using both tabGAP and soapGAP is similar as compared to 
DFT with somewhat 
larger variance in case of tabGAP, when compared to the soapGAP variance. However, the computational speed with the tabGAP is 20,000 and 400 times higher than the corresponding values for the DFT and soapGAP computations, respectively.}
Therefore, we mainly use the tabGAP in this work to carry out the sufficiently large number of TDE simulations. 
The details of the ML-IAPs can be found in our previous work~\cite{zhao2023complex}. 

Our simulations are set as follows. Firstly we create a simulation cell consisted of 10,240 atoms ($48.6\times49.5\times47.2$ Å$^3$).
In the cell, the periodic boundary conditions are applied in all directions to avoid the finite-size effect. 
\FD{To maintain overall the rectangular shape of the simulation cell, we align the $a$-axis with the \hkl[201] direction in the monoclinic unit cell of the $\beta$-\ce{Ga2O3} lattice (the $b$ and $c$ axes are left intact), which effectively converts the monoclinic unit cell into the orthorhombic one~\cite{zhao2024crystallization}.} 
Then the prepared system is equilibrated at 300 K and 0 bar pressure and used to simulate the TDE values. 
Two regions are separated to play different roles in our simulations.
An outer region of 4 \r A is kept at 300 K as a thermostat region and the atoms in the inner region follow a micro-canonical ($NVE$) ensemble where the recoil events happen. 
One atom (either Ga or O) in the inner region is chosen randomly as the PKA and given an initial velocity in a random direction finally. The initial velocity corresponds to a kinetic energy ($E_\mathrm{k}$) of 2 eV which is increased by 2 eV in each subsequent iteration until a stable FP is observed. 
Then the $E_\mathrm{k}$ of PKA is decreased by 1 eV and the same procedure is repeated {\HH to determine the final $E_\mathrm{d}$}. 
Each simulation is allowed to evolve for 6 ps employing the adaptive timestep approach to provide sufficient time for formation of a stable defect. 
The OVITO~\cite{ovitostukowski2009visualization} and VESTA~\cite{momma2011vesta} are used for the purpose of analyzing and visualizing the simulation results.

\subsection{Defect identification} \label{sec:defect_identify}

The FP is one of the most common defects in crystalline materials that are created under ion radiation.
\FD{The structure of a vacancy is straightforward, since it is just a lattice site with no atom in it. 
An interstitial, however, has more complex nature as an extra atom may occupy different sites between the lattice sites in materials.}
In $\beta$-\ce{Ga2O3}, there are two {\JL non-equivalent} Ga sites (usually labeled as Ga1 for tetrahedral and Ga2 for octahedral sites) and three nonequivalent O sites (labeled as O1 and O2 for three-bonded oxygen sites and O3 for four-bonded ones), \FD{which makes the defect analysis in all five atomic configurations of these sites highly non-trivial.} 
{\red The supplementary material Figure~S1} displays the five distinct lattice sites in detail highlighting also the bond geometry.

\begin{figure}[htbp!]
 \includegraphics[width=8.6 cm]{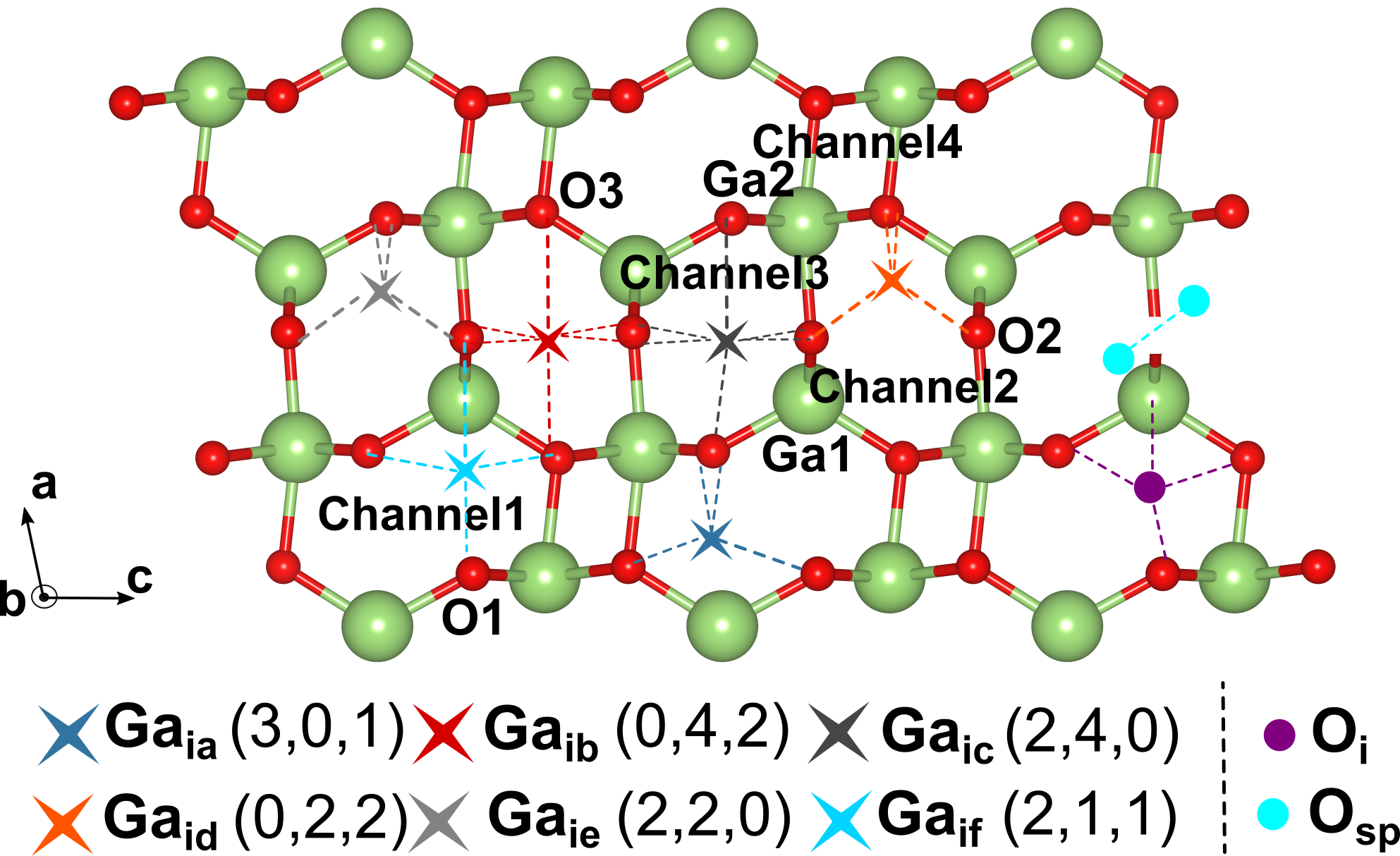}
 \caption{
  The possible Ga and O interstitial sites in $\beta$-\ce{Ga2O3}. 
  The ideal coordination numbers of Ga interstitial sites are shown in brackets in the legend. 
  Three numbers are the numbers of O1, O2 and O3 atoms, respectively, to which the site is bound. 
  The corresponding bonds are shown by the dotted lines. 
  The coordination numbers follows the notation introduced in Refs.~\cite{frodason2023migration, Oxygen_ingebrigtsenImpactProtonIrradiation2019}. 
  Green and red balls represent Ga and O atoms, respectively, consistent with the notation used below.}
 \label{fig:defectsites}
\end{figure}

As reported in previous studies \FD{in $\beta$-\ce{Ga2O3}~\cite{frodason2023migration, Gaint},} the six Ga interstitial ($\mathrm{Ga}_\mathrm{i}$) sites can be distributed in three different channels, \FD{as can be seen in Figure~\ref{fig:defectsites} ($\mathrm{Ga}_\mathrm{i}$ sites are shown as stars of different colors are distributed in three channels from Channel1 to Channel3)}. 
Besides, $\mathrm{Ga}_\mathrm{i}$ has recently been reported to prefer a split interstitial configuration as well~\cite{peelaersDeepAcceptorsTheir2019, frodason2023migration}. 
This is formed by two \ce{Ga} atoms in two adjacent channels that share one \ce{Ga} site in-between, which is different from those seen in traditional semiconductor materials. Hence, it is clear that the complex crystal system of $\beta$-\ce{Ga2O3} poses significant challenges in 
\FD{identifying the specific defects due to a variety of different atomic configurations for defective sites, which 
include split-sites as well, 
especially in over 5,000 simulations of the large atomic systems.}

One of the most commonly used techniques for identification of point defects produced in MD simulations of primary damage in crystals is the Wigner-Seitz (WS) analysis method~\cite{nordlund1998defect}.
Despite its computational efficiency and simplicity of application, the WS method fails to identify the defects in $\beta$-\ce{Ga2O3}, since neither it can distinguish between the different sites of $\mathrm{Ga}_\mathrm{i}$, nor it can detect the split interstitial. 
Hence, a more precise algorithm is required to identify the defects in this material. 

\FD{As it was mentioned above,} Figure~\ref{fig:defectsites} displays all six single $\mathrm{Ga}_\mathrm{i}$ sites ($\mathrm{Ga}_\mathrm{ia}$ to $\mathrm{Ga}_\mathrm{if}$), which are distributed in three channels, excluding the smallest rhombic one (Channel4). 
Among these sites, both $\mathrm{Ga}_\mathrm{ia}$ and $\mathrm{Ga}_\mathrm{if}$ are situated in the largest eight-edge channel (Channel1), while the remaining four $\mathrm{Ga}_\mathrm{i}$ sites are found in two similar six-edge channels.
Specifically, $\mathrm{Ga}_\mathrm{ib}$ and $\mathrm{Ga}_\mathrm{id}$ are found within the same channel (Channel2), which \FD{includes} O2 and O3 atoms. 
The interstitials of $\mathrm{Ga}_\mathrm{ic}$ and $\mathrm{Ga}_\mathrm{ie}$ are located in a different channel (Channel3) containing O1 and O2 atoms. 

\FD{To differentiate between the different $\mathrm{Ga}_\mathrm{i}$ sites,} we count the neighboring O atoms of each site within the radius 2.55 \r A.
This specific cutoff radius exactly corresponds to the distance between the first and second peaks of the Ga-O partial radial distribution function~\cite{zhao2023complex}.
The bonds that fall within the cutoff radius are shown for each site by the dashed lines of the corresponding to the site color in Figure~\ref{fig:defectsites}. 
Although the $\mathrm{Ga}_\mathrm{ia}$ and $\mathrm{Ga}_\mathrm{if}$ sites are bound only to O1 and O3 atoms in the largest channel, they are different by the situation of the site itself. 
While $\mathrm{Ga}_\mathrm{ia}$ site is located at the center of a tetrahedron, the $\mathrm{Ga}_\mathrm{if}$ site is closer to Ga2 atoms. 
\FD{Hence, following the notation introduced in Refs.~\cite{frodason2023migration, Oxygen_ingebrigtsenImpactProtonIrradiation2019} for description of the interstitial sites in $\beta$-\ce{Ga2O3}, the coordination number of the $\mathrm{Ga}_\mathrm{ia}$ site is identified as $(3,0,1)$, referring to} three O1 atoms and one O3 atom, while $\mathrm{Ga}_\mathrm{if}$ has a coordination number of $(2,1,1)$. 
\FD{Four sites in the six-edged channel (Channel3), $\mathrm{Ga}_\mathrm{ib}$, $\mathrm{Ga}_\mathrm{ic}$, $\mathrm{Ga}_\mathrm{id}$, and $\mathrm{Ga}_\mathrm{ie}$, correspond to two octahedral and two tetrahedral sites that are available in this channel.} 
\FD{While $\mathrm{Ga}_\mathrm{ib}$ and $\mathrm{Ga}_\mathrm{id}$ sites are bound to O2 and O3 atoms, they differ by the less O2 bonds in the $\mathrm{Ga}_\mathrm{id}$ tetrahedral site, hence,} their coordination numbers are $(0,4,2)$ and $(0,2,2)$, respectively. 
Similarly, $\mathrm{Ga}_\mathrm{ic}$ and $\mathrm{Ga}_\mathrm{ie}$ are bound to O1 and O2 atoms with less O2 atoms for the $\mathrm{Ga}_\mathrm{ie}$ tetrahedral site. 
Their respective coordination numbers are $(2,4,0)$ and $(2,2,0)$. 

Due to thermal effects and stochastic displacements during a PKA event, defects are not always found in the stable ideal sites as described above. 
Instead, they vibrate near the ideal sites, resulting in variations in their coordination numbers.
Therefore, additional criteria are also employed to identify and classify the specific defect type in order to better differentiate these sites. 
Details of the possible coordination number and other criteria can be found in {\red the supplementary material Table~S1 and Figure~S5}.

To enhance detection and classification of \ce{Ga}-related defects and their configurations in $\beta$-\ce{Ga2O3}, an optimized method combining WS analysis and coordination numbers has been developed. 
This method involves three steps:
\begin{itemize}
    \item Utilizing WS method, one $\mathrm{Ga}_\mathrm{i1}$ can be identified and its coordination number can be calculated to determine the specific type;
    \item All Ga atoms near the first $\mathrm{Ga}_\mathrm{i1}$, within the radius of 4 \r A are located. 
    Their coordination numbers are calculated separately to distinguish between normal sites of Ga atoms and $\mathrm{Ga}_\mathrm{i2}$.
    \item If no $\mathrm{Ga}_\mathrm{i2}$ is found, the $\mathrm{Ga}_\mathrm{i1}$ is regarded as a single Ga interstitial. 
    Otherwise, the split interstitial of $\mathrm{Ga}_\mathrm{i1-2}$ is identified.
\end{itemize}

With respect to oxygen interstitials, it is observed that the displaced O atom tends to form two types of interstitial defects: 
$\mathrm{O}_\mathrm{i}$ (single site) and $\mathrm{O}_\mathrm{sp}$ (O-O split interstitial), as reported in Ref.~\cite{Oxygen_ingebrigtsenImpactProtonIrradiation2019}, where the \textit{ab initio} simulations were carried out. {\HH As shown in Figure~\ref{fig:defectsites}},
$\mathrm{O}_\mathrm{i}$ is always located at the interstitial sites in the center of the channel, similar to $\mathrm{Ga}_\mathrm{i}$. While $\mathrm{O}_\mathrm{sp}$ is a O-O bonded {\JL defect}, {\HH where two oxygen atoms share one oxygen site}.
Based on our simulation results, another efficient method has been established to classify and analyze these two types of O interstitials. 
The classification of $\mathrm{O}_\mathrm{i}$ and $\mathrm{O}_\mathrm{sp}$ is based on the type of their shortest bond. 
If the shortest bond is between O and Ga (O-Ga bond), it is classified as $\mathrm{O}_\mathrm{i}$. 
Conversely, if the shortest bond is {\JL between two O atoms (O-O bond)}, it is categorized as $\mathrm{O}_\mathrm{sp}$. 
For more detailed information on the configurations of these defects see Section~\ref{sec:Ga&OFP}. 

{\HH In addition to identification of the defects, we also performed the calculations of the recombination energy barriers ($E_\mathrm{r}$) for these defects using the nudged elastic band (NEB) method. All of the energy barriers are calculated with 10,240 atoms and the total of 13 images are {\JL utilized} during each recombination path. 
The initial state is the system with a single FP and the final state is the perfect system. Additionally, all NEB simulations are stopped when the force is smaller than $5 \times 10^{-3}$ eV/\r A.}

\section{Results and discussion} 

\subsection{Quasi-static calculations}
\label{sec:quasistatic}

\FD{We performed a series of quasi-static calculations using the molecular static (MS) method~\cite{byggmastar2018effects, hamedani2021primary} in order to verify the reliability of our ML-IAPs in simulations of radiation damage to ensure the accuracy of the TDE values.} 
In the quasi-static calculations, one atom is moved towards a neighboring atom and the potential energies along this path are calculated with both the soapGAP and the tabGAP~\cite{zhao2023complex}, which can then be directly compared to single-point DFT calculations. This mimics the early stage of a recoil event but in a rigid well-defined lattice. All of the tested systems consisted of 160 atoms, allowing a comprehensive comparison with the DFT results. Detailed information regarding the DFT parameters used in these calculations can be found in {\red the supplementary material Appendix~C}.

\begin{figure*}[htbp!]
 \includegraphics[width=14 cm]{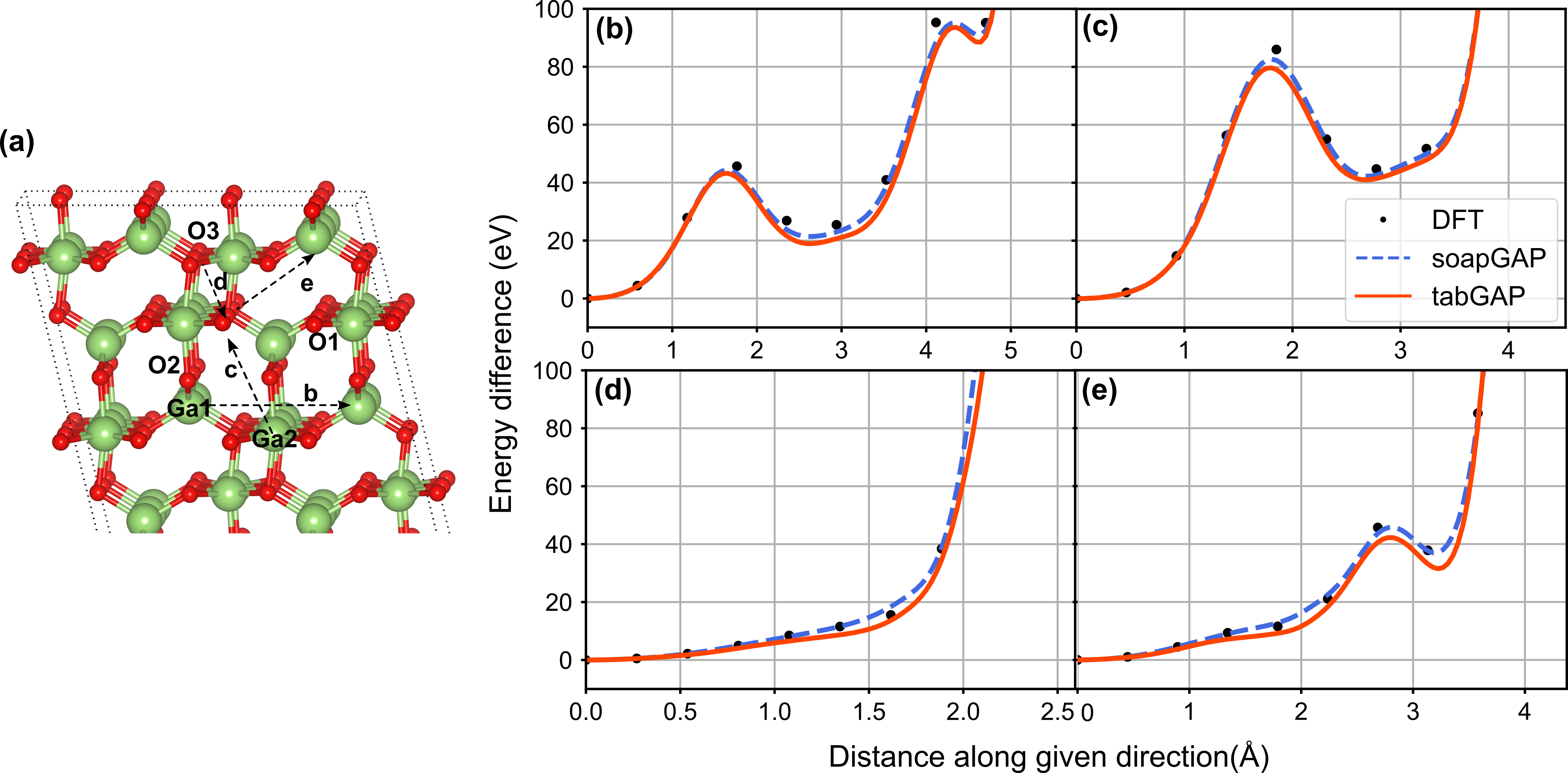}
 \caption{
 (a) Schematic representation of the four representative directions of atomic movement. Total energy difference for quasi-static simulations in \ce{Ga2O3} using soapGAP, tabGAP and DFT methods. (b) Ga1 $\rightarrow$ Ga1; (c) Ga2 $\rightarrow$ O3; (d) O3 $\rightarrow$ O3; (e) O3 $\rightarrow$ Ga1.
}
 \label{fig:quasi-static}
\end{figure*}

Four representative directions are calculated, as depicted in Figure~\ref{fig:quasi-static}(a) and the corresponding changes in total energy along these paths are shown in Figure~\ref{fig:quasi-static}(b)-(e).
Both curves predicted by the ML-IAPs exhibit good agreement with DFT data in describing the energy changes during the movement of one atom. 
This holds true even for more complex systems and shorter ranges. 
The largest discrepancy between the soapGAP and DFT is less than 18 meV/atom, which is small considering the fairly large values of the total energy change on the order of tens of eV for the whole system (hundreds meV/atom). 
Although tabGAP slightly underestimates the energy values at certain points compared with soapGAP and DFT, it still demonstrates a comparable accuracy and exhibits similar trends with soapGAP. 
It is important to note that our database does not include any charged point defects explicitly, so the observed differences fall within an acceptable range. 
These findings suggest that ML-IAPs can be effectively utilized for TDE simulations. 
We found similarly good agreement between the DFT and ML-IAP results for several other directions, which can be seen in {\red the supplementary material Figure~S2.}

\subsection{Threshold displacement energy} \label{sec:tdecalculations}

Figure~\ref{fig:defectsites} and {\red the supplementary material Figure~S1} both illustrate the coordination environments of the five sites in $\beta$-\ce{Ga2O3}. 
The Ga1 site exhibits a tetrahedral coordination with 1${\times}$O1, 2${\times}$O2 and 1${\times}$O3, while the Ga2 site shows octahedral coordination with 2${\times}$O1, 1${\times}$O2 and 3${\times}$O3. 
\FD{We note that the three O sites also have different coordination: O1 and O2 sites are 3-fold-coordinated, whereas the O3 is a 4-fold-coordinated site.}
Hence, \FD{to ensure that statistical properties of the $E_\mathrm{d}$ value for Ga and O in $\beta$-\ce{Ga2O3} was done correctly, }
it is necessary to \FD{distinguish between all five atomic types as described above.} 

\begin{figure}[htbp!]
 \includegraphics[width=8.6cm]{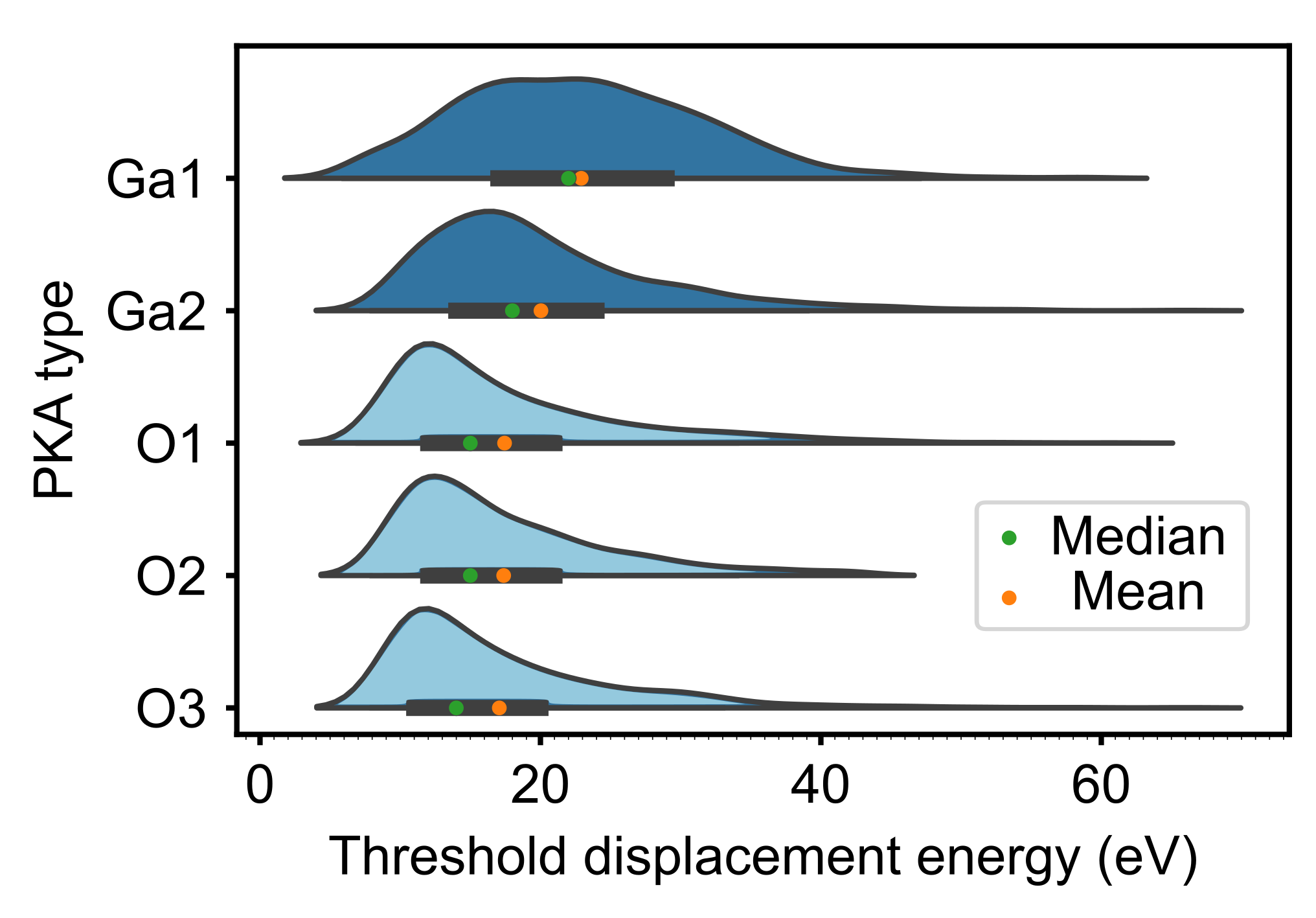}
 \caption{The distributions of $E_\mathrm{d}$ for the five PKA types in $\beta$-\ce{Ga2O3}. 
 The curves depict the counts and the black bars represent the $E_\mathrm{d}$ values ranging from 25\%-75\% of each PKA type. 
 The maximum, minimum, median and mean values are all included in the distributions.
 The statistics for each PKA type is over 1,000 random directions.}
 \label{fig:TDE-plot}
\end{figure}

To calculate the TDE values, we carried out the tabGAP MD simulations of different types of PKA in approximately 5,500 directions, as described in Section~\ref{sec:tde_method}. 
The resulting distributions of values for all types of PKAs are presented in Figure~\ref{fig:TDE-plot}, where each type of PKA is simulated over 1,000 random directions.

\FD{We average the TDE values over all directions for each atom site according to the formula}~\cite{nordlund2006molecular}:
\begin{equation}
    E_\mathrm{ave}= \frac{\iint E(\theta,\phi) \sin{(\theta)} \, \mathrm{d}\theta \,\mathrm{d}\phi}{\iint \sin{(\theta)} \,\mathrm{d}\theta \,\mathrm{d}\phi}
\label{equ:TDE}
\end{equation}

{\HH In this study, the polar angle ($\theta$) represents the angle between the velocity vector and the \hkl[010] direction, while the azimuthal angle ($\phi$) is the angle between the velocity vector projected on the \hkl(010) plane and \hkl[001] direction, where $\theta \in (0\degree, 180\degree)$ and $\phi \in (0\degree, 360\degree)$.} 
As shown in Figure~\ref{fig:TDE-plot}, the statistical analysis of $E_\mathrm{d}$ reveals distinct distributions for Ga and O PKAs. 
The mean and median $E_\mathrm{d}$ values for Ga1 are 22.91 eV and 22 eV, whereas those of Ga2 are 20.04 and 18 eV, respectively. 
Somewhat higher values of $E_\mathrm{d}$ for a Ga1 atom suggest that Ga1 is more resistant to {\JL the displacement} from its original lattice site. 
This difference in the mean $E_\mathrm{d}$ values can be attributed to their distinct local structures. The atomic configuration in {\red the supplementary material Figure~S1} illustrates that Ga1 occupies a tetrahedral site, whereas Ga2 adopts the structure of an octahedral site with longer bonds compared to Ga1-O ones.
\FD{Furthermore, we analyze the distribution of the maximum change in the total energy in quasi-static simulations (see Section~\ref{sec:quasistatic}), when we displaced Ga1 and Ga2 atoms step-by-step radially away from their original sites within 3~\r A. 
The distributions show systematic scans of all azimuthal directions with the interval of $1\degree$ around the corresponding sites on different crystallographic planes, see {\red the supplementary material Figure~S3}.}  
The comparison of these distributions reveals that the \FD{energy change for Ga2 in these calculations is much stronger than that for Ga1, especially on} \hkl(100) and \hkl(001) planes.
Consequently, the local atomic environment around Ga1 is denser, hence, the displacement of Ga1 energetically less favorable.

Moreover, the TDE of Ga1 is distributed broader over larger range of different values. 
It is also closer in shape to a Gaussian-type distribution, when compared to the distribution of TDE for Ga2, see the closer position of 
the mean and the median values for the two top distributions in Figure~\ref{fig:TDE-plot}. 
{\HH We explain this feature by the short-range symmetry of Ga1 sites. 
As shown in {\red the supplementary material Figure~S3(a-1) and (b-1)}, the behavior of Ga1 exhibits a nearly symmetric trend along the \hkl[201] direction, while the environment of Ga2 sites are more asymmetric due to the different structures of the neighboring Channel1 and Channel3.}

In contrast, despite the differences in the local environments of the three types of O PKAs, their $E_\mathrm{d}$ values are distributed very similarly, with only insignificant variation in the differences between the mean and the median values of these distributions. 
The mean $E_\mathrm{d}$ of O1, O2 and O3 are 17.44, 17.38 and 17.07 eV, while their median values are 15, 15 and 14 eV, respectively. 
The major difference for all three types of O atoms lies in the value of maximum $E_\mathrm{d}$, which exceeds 60 eV for both O1 and O3, while it only reaches 43 eV for O2 atoms 
(although our sampling is finite, Figure~\ref{fig:TDE-plot} clearly shows that values above 40 eV are statistically rare).

\begin{figure*}[htbp!]
 \includegraphics[width=16 cm]{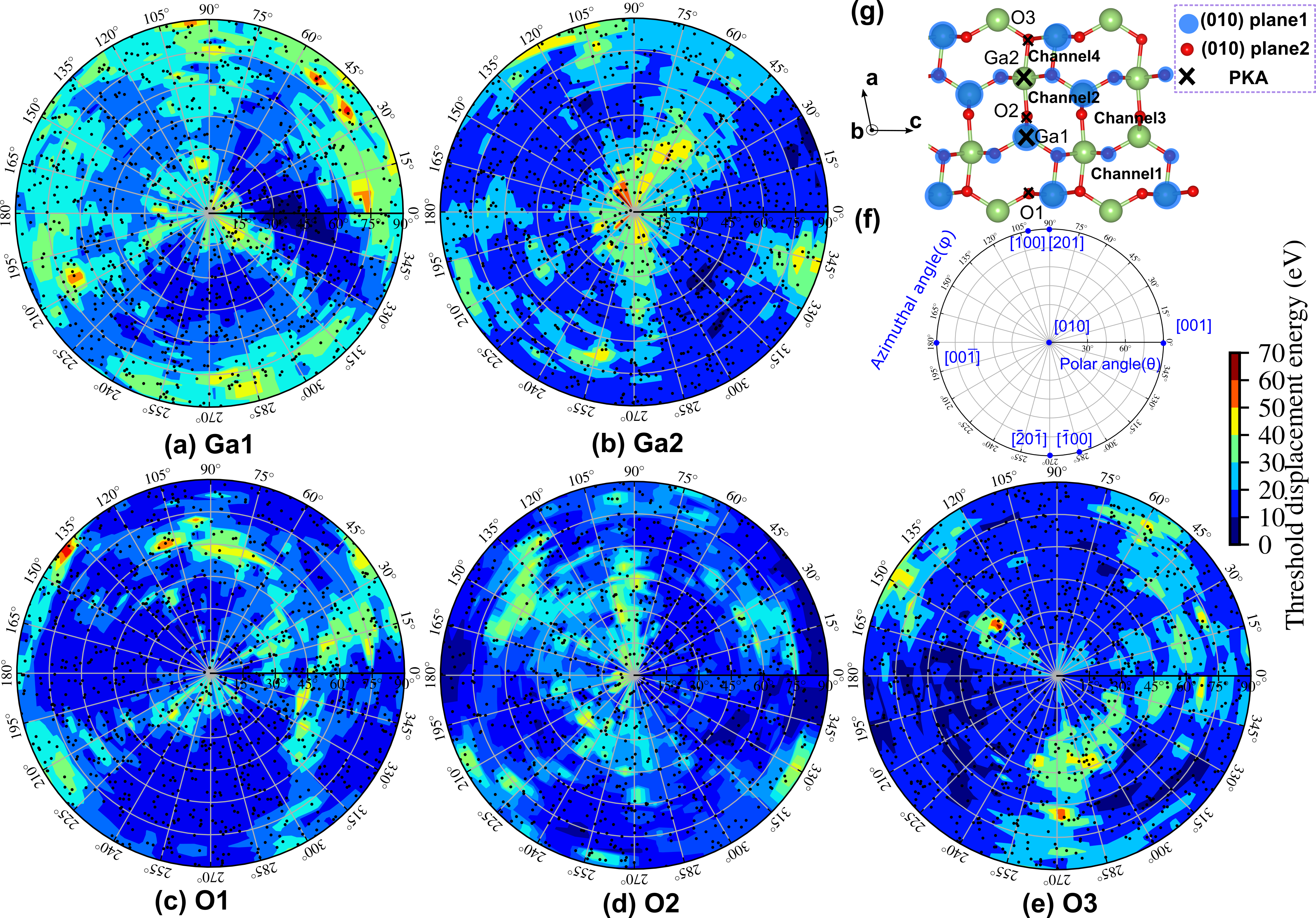}
 \caption{
 The separate TDE maps of (a) Ga1, (b) Ga2, (c) O1, (d) O2 and (e) O3 PKAs. 
 All calculated directions are included in the separate TDE maps and the maps are interpolated between the black dots with the nearest neighbor method. 
 (g) The crystal system and the studied PKAs from the viewpoint of the \hkl[010] direction. 
 The cross symbols label the studied PKAs and the blue transparent balls correspond to the atoms located on a same \hkl(010) plane while the green or red atoms are on another \hkl(010) plane.  
 (f) Diagram of the angles used in the TDE maps and some pinpointed low Miller-index directions shown as blue dots.
} 
 \label{fig:TDE-map}
\end{figure*}

Considering the significant directional dependence of $E_\mathrm{d}$ values as demonstrated in Refs.~\cite{nordlund2006molecular,nordlundLargeFractionCrystal2016}, we further generate the 
TDE maps over all directions scanning the polar ($\theta$) and azimuthal ($\phi$) 
taking into account the {\JL symmetries} of the monoclinic Bravais lattice of the $\beta$-\ce{Ga2O3}. 
For comprehensive maps we performed the transformations of two types:

\begin{itemize}
\item $\theta$ is firstly reduced to $\in (0\degree, 90\degree)$ and $\phi$ remains $\in (0\degree, 360\degree)$ according to the symmetry of the \hkl(010) plane. 
It is important to note that this is the only form of symmetry for all five sites:
 \begin{equation}
\theta = \begin{cases}
\theta, & \text{if } \theta \leq 90 \\
180 - \theta, & \text{if } \theta > 90
\end{cases}
\label{equ:TDE2}
\end{equation}
\item Each atomic site is transformed to the studied PKAs which are shown in Figure~\ref{fig:TDE-map}(g), since for any atomic site there are two symmetrical atoms in $\beta$-\ce{Ga2O3}. 
 \begin{equation}
\phi = \begin{cases}
\phi, & \text{if atom $\neq$ the studied PKA}  \\
180 + \phi, & \text{if atom $\equiv$ the studied PKA} 
\end{cases}
\label{equ:TDE3}
\end{equation}
\end{itemize}

{\HH Based on this transformation, the overall TDE maps, as shown in {\red the supplementary material Figure~S6}, illustrates that the central region corresponding to the \hkl[010] direction exhibit higher $E_\mathrm{d}$ values compared to {\JL the rest of the map}. This can be attributed to the lattice structure of $\beta$-\ce{Ga2O3}, where atoms of the same type are predominantly positioned along the \hkl[010] direction, resulting {\JL in a large short-distance repulsion for PKAs to displace from their original sites along this direction.} Since the atomic environment around different atomic types are {\JL highly diverse}, five separate TDE maps divided by atomic types are shown in Figure~\ref{fig:TDE-map}(a)-(e).}

As depicted in Figure~\ref{fig:TDE-map}(a) and (b), Ga1 and Ga2 PKAs exhibit distinct patterns. For Ga1 PKA, directions with $\theta$ near $90\degree$ have higher $E_\mathrm{d}$ values with most of them exceeding 30 eV. 
All $E_\mathrm{d}$ values larger than 60 eV are in these directions, {\JL indicating} that \FD{the Ga1 PKA is the least likely to be displaced in the direction almost parallel to the \hkl(010) plane.}
With respect to the minimum $E_\mathrm{d}$, the TDE map shows that the Ga1 PKA is the easiest to be displaced along these two series of directions, one is $\theta \in (30\degree, 60\degree)$ and $\phi \in (-15\degree, 15\degree)$, and another one is $\theta \in (22\degree, 45\degree)$ and $\phi \in (210\degree, 300\degree)$. 
These two series of directions both show $E_\mathrm{d}$ values below 20 eV. 
Analysing the structure, we find that the former series corresponds exactly to the directions towards the lower part of the Channel2 and the latter one corresponds to the directions through the largest eight-edge channel (Channel1). 
For Ga2 PKA, we observe that most directions with $\theta$ near $90\degree$ provide an easier pathway for displacement from the original sites, except when $\phi$ is near $110\degree$ and $345\degree$. 
The maximum $E_\mathrm{d}$ for Ga2 is associated with the directions around \hkl[010], suggesting that Ga2 follows a different pattern from Ga1. 
For the minimum values, there are three blue-colored regions on the TDE map.
These three regions represent the two largest channels (Channel1) on the two sides [$\phi \in (0\degree, 30\degree)$ and $\phi \in (120\degree, 165\degree)$, $\theta \in (30\degree, 90\degree)$] and the Channel2 [$\phi \in (285\degree, 330\degree)$, $\theta\in (30\degree, 90\degree)$].
In addition, the directions corresponding to the minimum values for Ga1 and Ga2 PKAs indicate their tendency to enter into the Channel1 and Channel2, though they have different coordination system. 

Despite the fact that the three O atoms have similar mean $E_\mathrm{d}$ from the data in Figure~\ref{fig:TDE-plot}, their TDE maps exhibit different directional dependencies. 
As shown in Figure~\ref{fig:TDE-map}(c), O1 demonstrates a preferred direction for displacement when $\phi \in (120\degree, 315 \degree)$, which corresponds to the directions towards the neighboring Channel1 and Channel3. 
Along these directions, {\HH there are nearly no atoms located in proximity to O1 and the two exceptional regions precisely correspond to the positions of the two Ga atoms located in the same Channel1 as O1, and on the same \hkl(010) plane, as displayed in Figure~\ref{fig:TDE-map}(g). 
One of the Ga atoms is the non-bonded Ga2 atom, which is located in the range of $\phi \in (135\degree, 150\degree)$, while another one is the bonded Ga1 located in the range of $\phi \in (195\degree, 225\degree)$ [Figure~\ref{fig:TDE-map}(c)].}
Considering the fact that Ga atoms are larger and heavier, O1 PKAs are indeed difficult to displace along these directions. 
In addition, in regions where $\phi$ is less than $120\degree$, relatively high $E_\mathrm{d}$ values are observed compared to other areas. This could be attributed to the presence of bonded Ga2 atom and Ga1 atom located in the same Channel1 as O1, {\JL and hence}, the O1 PKAs tend to move into the Channel1 and Channel3.

For O2 PKAs, the structure demonstrates a higher degree of symmetry between the two adjacent six-edge channels, Channel1 and Channel2, compared with the other two O atoms, which is also supported by the data shown in {\red the supplementary material Figure~S4(b)}.
Therefore, it is anticipated to demonstrate symmetrical directions on its TDE map. 
As depicted in Figure~\ref{fig:TDE-map}(d), the minimum $E_\mathrm{d}$ is distributed along two symmetrical directions, \hkl[001] and \hkl[00-1], indicating that the O2 PKAs are most easily displaced in these two directions.
Notably, only one O2 atom is present along these directions. In addition, two symmetrical regions [$\theta \in (75\degree, 90\degree)$] where $\phi \in (315\degree, 345\degree)$ and $\phi \in (195\degree, 225\degree)$ also show that O2 PKAs are difficult to displace along these directions. 
These symmetrical directions correspond to the nearby Ga2 atoms on two sides. 
However, we also observe that not the whole TDE map displays the complete symmetry for O2 PKA. 
The left part of the TDE map exhibits a clearly brighter distribution than the right part, particularly for the top part. 
Therefore, O2 PKA shows a preference for moving into Channel2, rather than Channel3.

With respect to O3 PKAs, almost the whole region where $ \phi \in (165\degree, 245\degree)$ 
in Figure~\ref{fig:TDE-map}(e) show rather low $E_\mathrm{d}$ values, no matter what $\theta$ is. This large region with small $E_\mathrm{d}$ values is exactly {\JL confined} by the bonds of O3-Ga1 and O3-Ga2 on the same \hkl(010) plane. 
Another region exhibiting low $E_\mathrm{d}$ values is in the range of $\phi \in (75\degree, 120\degree)$, {\HH which corresponds to the adjacent Channel2. Along these directions, the repulsive force is weaker.}
Additionally, it can be found that the directions toward the nearby Ga atoms all have higher $E_\mathrm{d}$ values that exceed 30 eV. 
The Ga atom prevents the O3 PKA to be displaced into the Channel3.
However, the directions where $\phi \in (285\degree, 330\degree)$ is an exception, which means O3 PKA has a high probability to displace into the smallest channel. 
This tendency is different from the other two O PKAs since they do not present a high possibility to move into the Channel4, while the simulations of O3 PKA show it.

\subsection{FP configurations} \label{sec:Ga&OFP}
{\JL The predominant outcome following recoil events is the formation of FPs. 
Thus, we analyze the defect structures at 6 ps from the aforementioned 5,500 TDE simulations to explore the characteristics of radiation-induced defects.} 
{\JL For} all TDE simulations, {\JL a single FP is the predominant outcome, whereas} only 2.6\%  of the simulations involve {\JL the formation of double} FPs, primarily attributed to Ga PKA {\JL events}. 
Notably, these double FPs mainly consist of {\JL double Ga FPs (33.6\%) 
and mixed Ga/O FPs (64.2\%), with double O FPs being infrequent (2.2\%).}
{\JL Moreover, the total number of O-related defects surpasses that of O PKAs, considering all simulations, whereas the opposite trend is observed for Ga-related defects and Ga PKAs. 
These findings corroborate the TDE results, emphasizing that oxygen atoms are more susceptible to displacement than Ga atoms.}

\begin{figure*}[htbp!]
 \includegraphics[width=16 cm]{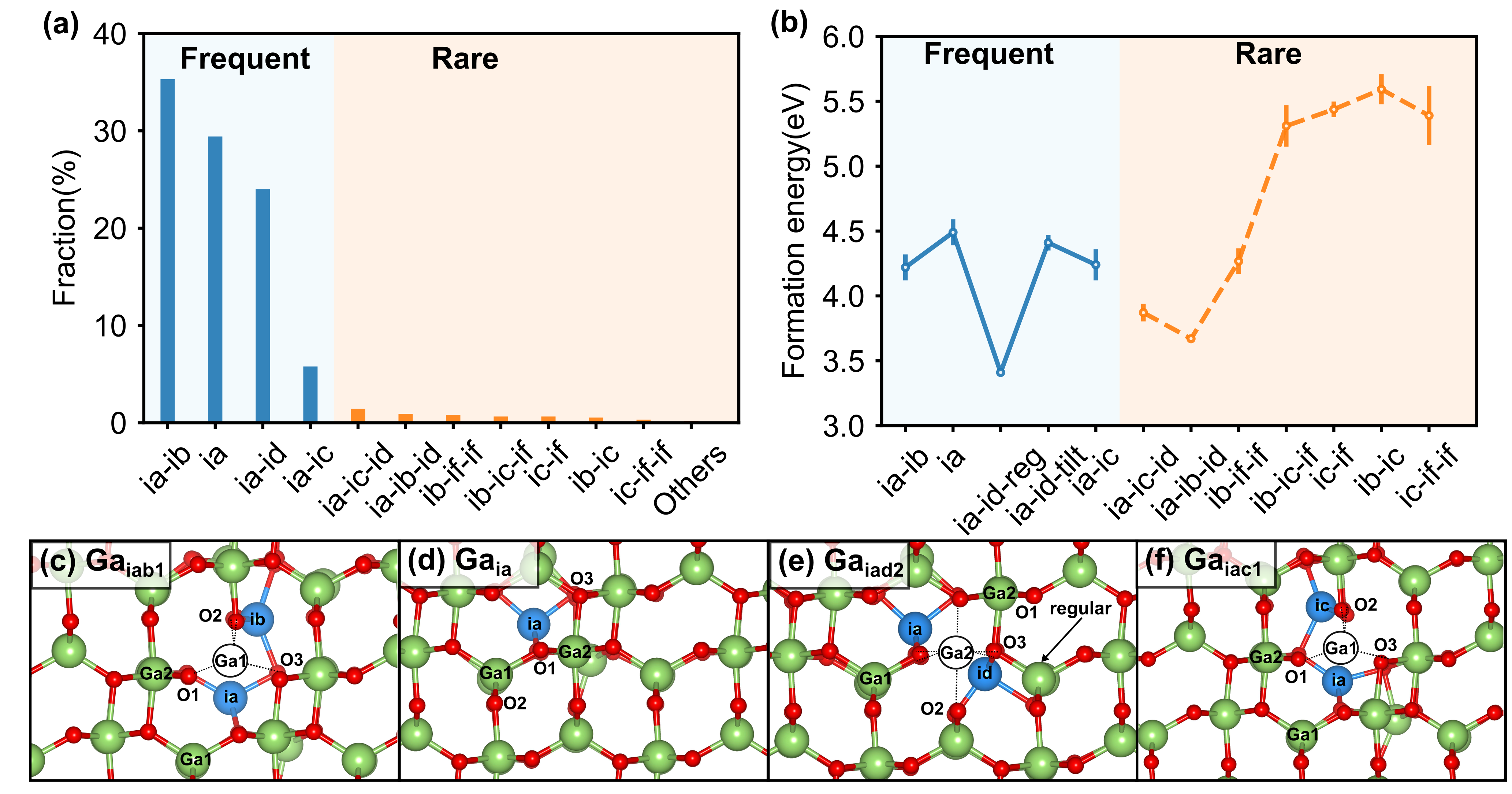}
 \caption{(a) The number of different Ga FP types at 6 ps of TDE simulations. Ga FPs that have a fraction higher (lower) than 5\% are categorized as frequent (rare) configurations. The total number of 1,866 Ga FPs are analyzed in this study. (b) Formation energy of different types of Ga FP from TDE results after structure relaxation. 
 {\JL The error bars show the standard variances from the different locations of the Ga vacancy.}
 (c)-(f) Configurations of the interstitial in the four frequent FP configurations. 
 Blue atoms represent the interstitials and white atoms symbolize the shared sites between Ga split-interstitials. 
 $\rm V_{Ga}$ are not shown here. 
 The names of the FPs are represented by their interstitial type as presented in Figure~\ref{fig:defectsites}.}
 \label{fig:Gai_num_EF}
\end{figure*}

{\JL In Figure~\ref{fig:Gai_num_EF}, we initiate our analysis of Ga FP by examining the outcomes of Ga interstitial ($\rm Ga_{i}$) structures resulting from 1,700 TDE simulations.} 
{\JL Given the extensive range of possible Ga vacancy sites, we exclusively categorize the $\rm Ga_{i}$ structures of the Ga FPs.} 
Differentiating by the structures of $\rm Ga_{i}$, more than ten types of Ga FPs are identified. 
{\JL The individual proportions} are shown in Figure~\ref{fig:Gai_num_EF}(a), where four configurations -- ia-ib, ia, ia-id and ia-ic -- account for 95\% of the total. 
Due to their higher frequency of occurrence, these configurations are designated as frequent. 
{\JL On the other hand, additional structures appear with smaller quantities and classified as rare configurations.} 
Formation energies ($E_\mathrm{f}$) of these configurations after relaxation with the molecular static method are calculated as described in Refs.~\cite{frodason2023migration,heDynamicsStudiesNitrogen2020}: 
\begin{equation}
    E_\mathrm{f}= {E_\mathrm{FP}}-{E_\mathrm{bulk}},
\label{equ:formation_energy}
\end{equation}
where $E_\mathrm{FP}$ is the total energy of the system with the FP while $E_\mathrm{bulk}$ corresponds to the bulk crystal without any defects. The related results are depicted in Figure~\ref{fig:Gai_num_EF}(b). {\JL Notably, disparities between the proportions and $E_\mathrm{f}$ values are observed, \textit{i.e.}, the most populated configuration does not exhibit the lowest $E_\mathrm{f}$, suggest a kinetic effect rather than a purely thermodynamic trend. Consequently, for a more profound understanding of Ga FP structures, the subsequent discussion delves into their atomic structures in detail, as presented in Figure~\ref{fig:Gai_num_EF}(c)-(f).}

As the most populated one, $\rm Ga_{iab1}$ is a split-interstitial combining the sites of ia and ib, as shown in Figure~\ref{fig:Gai_num_EF}(c). 
$\rm Ga_{ia}$ and $\rm Ga_{ib}$ are located in different channels and share one normal Ga1 site. 
This FP has a low $E_\mathrm{f}$ of $4.22\pm0.01$ eV. 
It is also worth noting that these two interstitial sites, $\rm Ga_{ia}$ and $\rm Ga_{ib}$ are on the same \hkl(010) plane. 
The second most common Ga FP is $\rm Ga_{ia}$, which is also the only interstitial with a single site. 
From Figure~\ref{fig:Gai_num_EF}(d), it induces a large distortion of nearby atoms and the $E_\mathrm{f}$ of it is approximately $4.49\pm0.1$ eV. 

{\JL Figure~\ref{fig:Gai_num_EF}(e) displays} the structure of $\rm Ga_{iad2}$, which is similar to $\rm Ga_{iab1}$. 
Though $\rm Ga_{id}$ and $\rm Ga_{ib}$ are both in the same channel, their sharing site is different. 
Only Ga2 sites provide the possibility to form $\rm Ga_{iad2}$ while Ga1 sites form $\rm Ga_{iab1}$. 
Therefore, it is also easier to distinguish $\rm Ga_{iab1}$ and $\rm Ga_{iad2}$ according to the atomic type of their sharing Ga sites. 
Furthermore, the calculations of $E_\mathrm{f}$ also reveal that $\rm Ga_{iad2}$ has two distinct types, $\rm Ga_{iad2-reg}$ (regular) and $\rm Ga_{iad2-tilt}$. $\rm Ga_{iad2-reg}$ has the lowest $E_\mathrm{f}$ ($3.41\pm0.01$ eV) among all Ga FPs while that of $\rm Ga_{iad2-tilt}$ is $4.41\pm0.06$ eV. 
Analysing their structure, the $\rm Ga_{iad2-reg}$ FP is a special structure since the vacancy is always at the Ga1 site adjacent to $\rm Ga_{id}$, as the arrow shows in Figure~\ref{fig:Gai_num_EF}(d). 
In this case, the $\rm Ga_{id}$ interstitial binds with two neighboring O2 atoms equally due to two symmetrical vacant sites around it. 
In contrast, the $\rm Ga_{id}$ of the $\rm Ga_{iad2-tilt}$ FP forms uneven bonds due to the vacancy being located at other positions. 
Furthermore, by using the NEB method, the recombination barrier of $\rm Ga_{iad2-reg}$ is calculated. The results show that the barrier is very low, only 0.26 eV. 
Such a low barrier reveals that though $\rm Ga_{iad2-reg}$ is the FP with lowest formation energy, it is easier to be recombined during the radiation process and hence not the most frequent configuration.

Lastly, the fourth stable Ga FP is $\rm Ga_{iac1}$, as displayed in Figure~\ref{fig:Gai_num_EF}(d). 
Different from the above three configurations, this is the only stable FP in which one of interstitial sites (ic) is located within the channel formed by O1 and O2 (Channel3). 
The formation energy of this FP is calculated to be $4.24\pm0.12$ eV.
However, the frequency of occurrence is not as high as that of $\rm Ga_{iab1}$, even though the $E_\mathrm{f}$ is similar, implying that formation energy is not the only factor to determine the occurrence of them. 

In addition to the above four stable configurations, we also observed some more complex structures from the TDE simulations, \textit{i.e.}, $\rm Ga_{ibff}$ and $\rm Ga_{ibcf}$. 
They are also stable after relaxation and the majority of them are three interstitial sites sharing two vacancies. 
However, their production probabilities are very low. 
Hence, they are classified as rare configurations. 
In addition, they are easier to be transformed into the above stable configurations. 
For instance, $\rm Ga_{ibff}$ exhibits a transformation barrier of 0.22 eV to $\rm Ga_{iab1}$. 
Detailed information on these rare structures can be found in {\red the supplementary material Figure~S7}.

\begin{figure*}[htbp!]
 \includegraphics[width=16 cm]{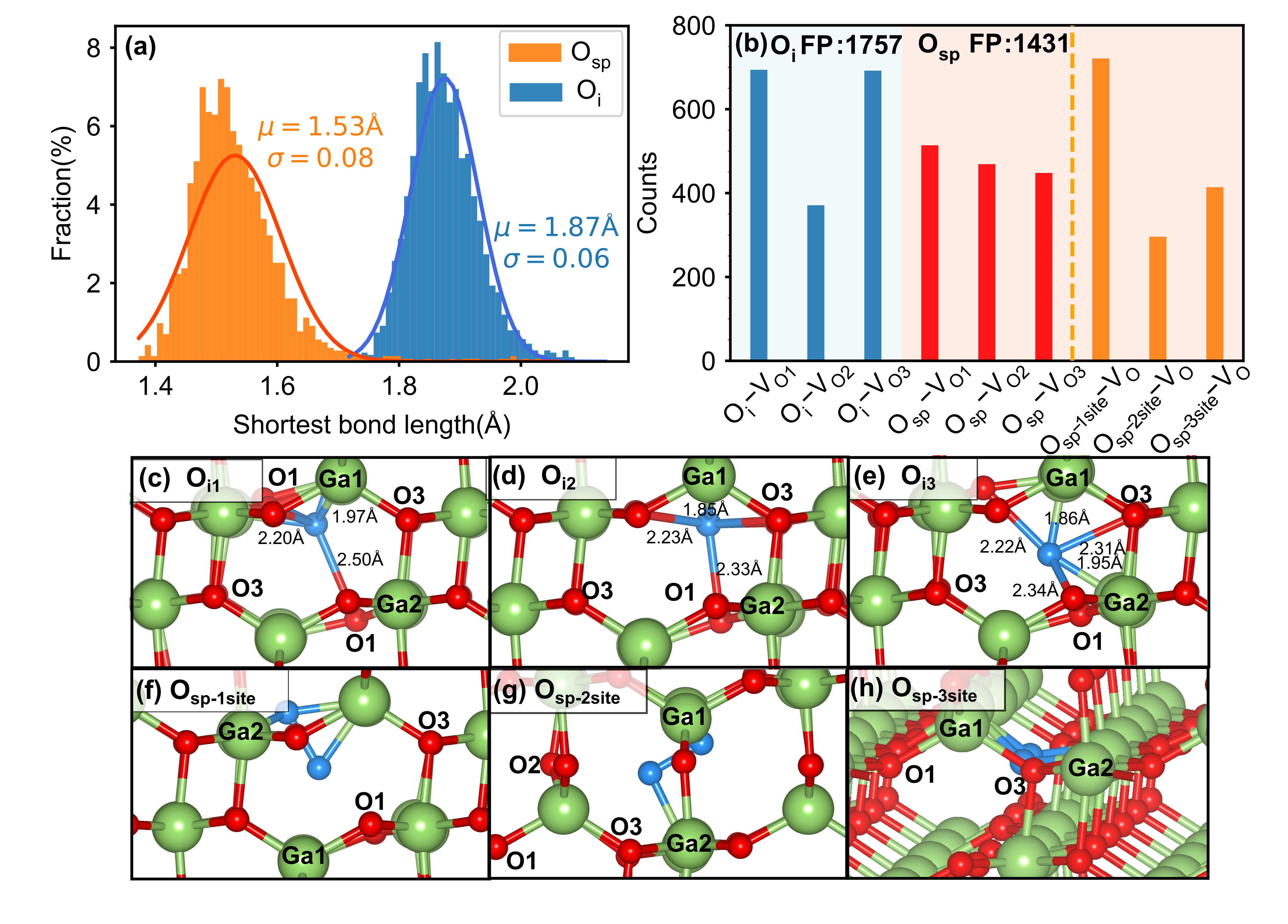}
 \caption{(a) The distribution of the bond length of the interstitial with its nearest atom in $\rm O_{i}$ and $\rm O_{sp}$ FPs.  (b) The number of different types of O FPs separated by the three O atomic sites at 6 ps of TDE simulations. {\HH $\rm O_{i}$ FPs are only separated by the sites of the vacancy ($\rm O_{i}$-$\rm V_{Oi}$, blue bars). $\rm O_{sp}$ FPs are separated by the sites of the vacancy ($\rm O_{sp}$-$\rm V_{Oi}$, red bars) and interstitial ($\rm O_{sp\text{-}isite}$-$\rm V_{O}$, orange bars), respectively}. $\rm V_{Oi}$ corresponds to the vacancy sites on O1, O2 and O3 atomic types and $\rm O_{sp\text{-}isite}$ is the split-interstitial on three O atomic types. (c)-(e) $\rm O_{i}$ and (f)-(h) $\rm O_{sp}$ FP configurations created by the low-energy radiation. Blue balls represent the O interstitial defects ($\rm V_{O}$ is not shown in these diagrams).} 
 \label{fig:O-FP_length&num}
\end{figure*}

Due to the smaller size of O atoms, O FPs are easier to be produced than Ga FPs as discussed above. 
Unlike the anion defects in GaN~\cite{heDynamicsStudiesNitrogen2020} or SiC~\cite{Cint-PhysRevB.69.235202}, two distinct types of O FP are produced during TDE simulations. Distinguished by the bond type of O interstitial with its nearest atom,  $\rm O_{i}$ and $\rm O_{sp}$ are identified.
The shortest bond of O interstitial in $\rm O_{i}$ FP is an O-Ga bond while that of $\rm O_{sp}$ is an O-O bond. 
In addition, the lengths of their shortest bond are totally different, as displayed in the distributions in Figure~\ref{fig:O-FP_length&num}(a), which both exhibit a high level of statistical significance. 
With respect to $\rm O_{sp}$, the mean O-O bond of it is 1.53 \r A, which is a bit longer than oxygen molecules {\JL (1.20 \r A for \ce{O2})}. 
However, the lengths of the O-Ga bonds in $\rm O_{i}$ FPs are distributed mainly around 1.87 \r A, which is shorter than the length of normal O-Ga bond in $\beta$-\ce{Ga2O3}. 
The length of these two types bonds indicate that these two types of O FPs are stable. 

{\HH Since there are three types of oxygen atomic sites in $\beta$-\ce{Ga2O3}, based on the analysis of shortest bond type, the correlation between O ($\rm O_{i}$ and $\rm O_{sp}$) FPs and the types of oxygen atom sites is investigated further.} 
To explain the mechanism behind it, we also relax some structures after the above TDE simulations to obtain their stable configurations and calculated the formation energies of them. 
With respect to $\rm O_{i}$ FPs, we find the number of them is not related to the atomic type of the interstitial. 
By using the molecular statics method to relax the observed $\rm O_{i}$ FPs, three different configurations of $\rm O_{i}$ which are all located in the largest channel are observed. 
As illustrated in Figure~\ref{fig:O-FP_length&num}(f)-(h), $\rm O_{i1}$, $\rm O_{i2}$ and $\rm O_{i3}$ all have a high symmetry with its nearby atoms. 
$\rm O_{i1}$ and $\rm O_{i2}$ are both formed by two even O-Ga bonds with Ga1 atoms that are in the same \hkl[010] direction. However, $\rm O_{i1}$ is closer to the Ga1-O1 bond and only has two bonds with O1 atoms while $\rm O_{i1}$ has four extra equal bonds with O1 and O3 atoms, which is similar to $\rm Ga_{if}$ configurations. 
In contrast, $\rm O_{i3}$ is found to be connected with one Ga1 and one Ga2 atom and their bond lengths are not equal (1.86 \r A and 1.95 \r A). 
The $\rm O_{i3}$-Ga and $\rm O_{i3}$-O3 bonds are both parallel to the \hkl(010) plane which means this type of $\rm O_{i}$ is on the same plane with the bonded Ga1, Ga2 and O3 atoms. 
Even though the local bonding geometries of these three $\rm O_{i}$ are different, FPs with them are regarded as equivalent since they have nearly the same formation energies and are located in the same channel. 

Therefore, we further identify them according to the atomic type of the vacancy regardless of the exact $\rm O_{i}$ configuration. 
The number of them are shown in Figure~\ref{fig:O-FP_length&num}(b). 
It is clear that although almost 1,700 $\rm O_{i}$ FPs are produced during the TDE simulations, the fraction of $\rm O_{i}$-$\rm V_{O2}$ is much lower than that of the other two types, only half as frequent as those with $\rm V_{O1}$ and $\rm V_{O3}$. 
This large difference indicates that the O2 atom is more difficult to be displaced from its original sites to form $\rm O_{i}$ FPs. 
Analyzing the structure of the three different $\rm O_{i}$, we ascribe the reason for this to the formation mechanism. 
As discussed above, the structures of the three $\rm O_{i}$ are all located in Channel1 which is only surrounded by O1 and O3 atomic types.
Since O2 is far away from Channel1, O2 atoms cannot easily enter it and form $\rm O_{i}$ configurations.

In the case of $\rm O_{sp}$ FPs, they exhibit an opposite correlation.
As the red bars show in Figure~\ref{fig:O-FP_length&num}(b), the number of $\rm O_{sp}$ FPs with the three types of $\rm V_{O}$ are all about 450, which illustrates that they are independent of the vacancy site. 
However, the numbers of them differ a lot when divided by the specific atomic type of the interstitial. 
From the orange bars in Figure~\ref{fig:O-FP_length&num}(b), $\rm O_{sp\text{-}1site}$ FPs are much more frequent than the other two sites, produced about 700 times which is almost half of all $\rm O_{sp}$ FPs.
Therefore, it can be concluded that the most favoured $\rm O_{sp}$ configuration is an O-O bond at the O1 site. 

Furthermore, $E_\mathrm{f}$ of them at different sites are also calculated and compared after relaxing the structures. Among these three sites, {\HH $\rm O_{sp\text{-}2site}$ FPs exhibit the highest $E_\mathrm{f}$, measuring $5.07\pm0.03$ eV while $E_\mathrm{f}$ of $\rm O_{sp\text{-}1site}$ FPs and $\rm O_{sp\text{-}3site}$ FPs are only $4.18\pm0.02$ eV and $4.36\pm0.03$ eV, separately. }
Detailed atomic configurations of $\rm O_{sp}$ at the three sites after relaxation are presented in Figure~\ref{fig:O-FP_length&num}(f)-(h).
These configurations reveal distinct structural characteristics.
Although $\rm O_{sp\text{-}1site}$ and $\rm O_{sp\text{-}2site}$ have different orientations, they both lie parallel to the \hkl(010) plane.
In contrast, $\rm O_{sp\text{-}3site}$ tends to form a bond along the \hkl[010] direction.
The inconsistency of their structure likely stem from their coordination system.
{\HH In $\beta$-\ce{Ga2O3}, O1 and O2 both exhibit a coordination number of three Ga atoms while O3 is the only one with four-fold coordinated site.}

To ensure the reliability of the results, $E_\mathrm{f}$ of these three O FPs on different sites are also calculated with 160 atoms in DFT simulations.
While slight differences are observed between MD and DFT, the trends in $E_\mathrm{f}$ across different sites and the bond lengths exhibit good agreement. 
{\red The supplementary material Figure~S8} provides the detailed information on the comparable results obtained from MD and DFT calculations.

Additionally, Figure~\ref{fig:O-FP_length&num}(b) shows the amount of $\rm O_{i}$ and $\rm O_{sp}$ with the occurrence of 55\% and 45\% of all simulations. {\JL $E_\mathrm{f}$ of $\rm O_{i}$ FPs} is calculated as $4.29\pm0.03$ eV, while $E_\mathrm{f}$ of $\rm O_{sp}$ FPs is $4.55\pm0.04$ eV. 
The results of probability and $E_\mathrm{f}$ both demonstrate that $\rm O_{i}$ FPs are a bit easier to produce. 
However, regarding the specific types discussed above, $\rm O_{sp\text{-}1site}$ FP is produced more than 700 times, which is the highest occurrence among the six types of O FPs.

\subsection{Recombination behaviour} \label{sec:FP_recomb}

\begin{figure*}[htbp!]
 \includegraphics[width=16 cm]{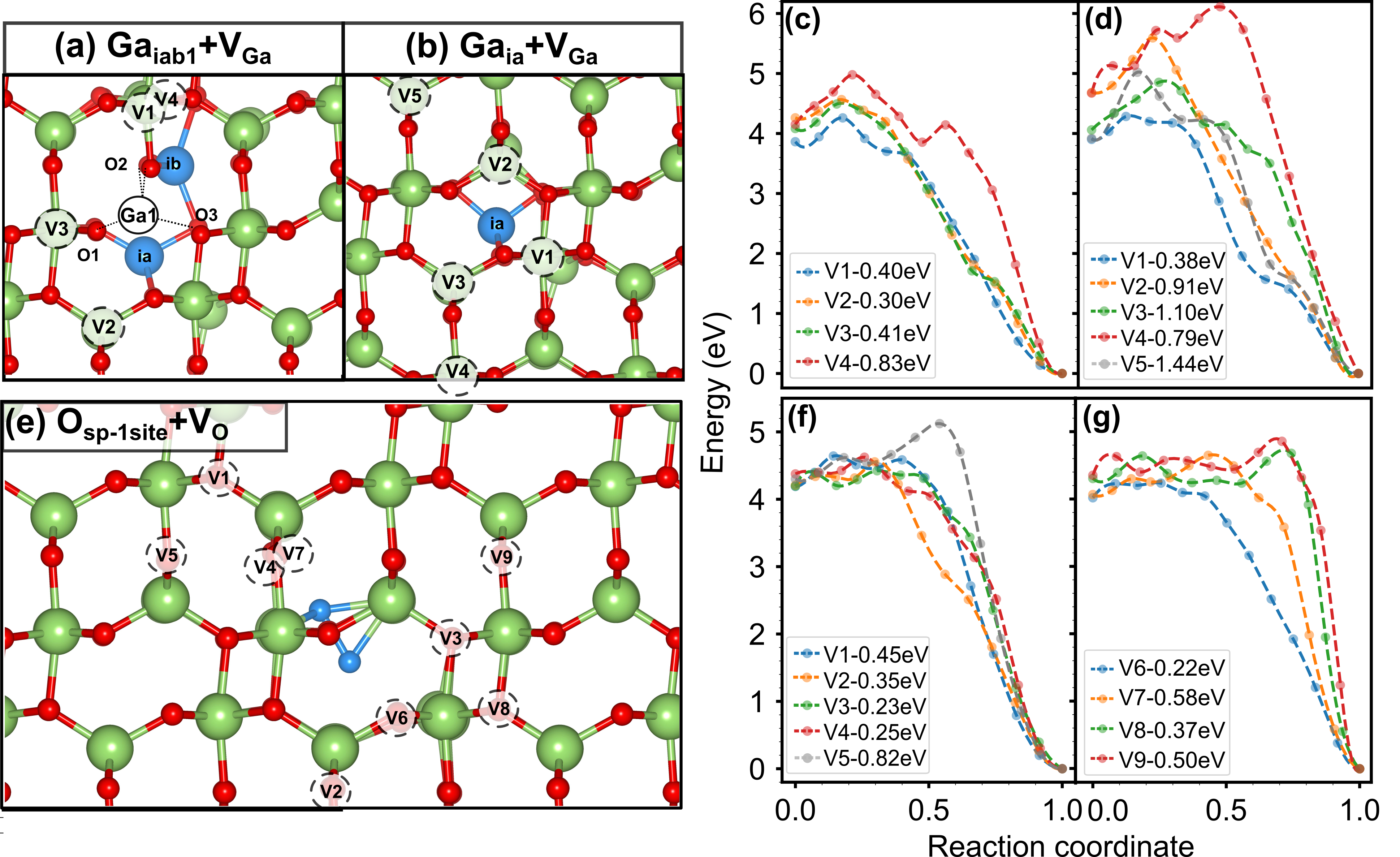}
 \caption{{\HH Left panel: The configurations of (a) four types of $\rm Ga_{iab1}$-$\rm V_{Ga}$, (b) five types of $\rm Ga_{ia}$-$\rm V_{Ga}$ and (e) nine types of $\rm O_{sp\text{-}1site}$-$\rm V_{O}$. Right panel: The corresponding path and $E_\mathrm{r}$ for recombination behaviour of (c) $\rm Ga_{iab1}$-$\rm V_{Ga}$, (d) $\rm Ga_{ia}$-$\rm V_{Ga}$ and (f)(g) $\rm O_{sp\text{-}1site}$-$\rm V_{O}$. The blue balls are the sites of the interstitials and the white dashed balls are the sites of vacancies. The numeric labeling of vacancies is arranged in order of increasing distance between them and the interstitial. All of the solid points are results calculated with the NEB method with 10,240 atoms and the dashed lines are interpolated using cubic splines. The first point is the energy of the system with the given FP and the last point refers to the perfect system with no defects.}}
 \label{fig:Ga&O-FP_recomb}
\end{figure*}

The recombination behavior of FP often plays a significant role in the recovery processes of irradiated semiconductor materials.
Therefore, the recombination behaviours of Ga/O FPs produced in our TDE simulations are investigated {\JL using the NEB method}. Since there are a variety of FP types produced in our TDE simulations, we mainly focus on the types that are most frequent based on the above analysis. 
For Ga FPs, two types of $\rm Ga_{iab1}$-$\rm V_{Ga}$ and $\rm Ga_{ia}$-$\rm V_{Ga}$ are investigated, while $\rm O_{sp\text{-}1site}$-$\rm V_{O}$ is considered for O FPs. The configurations and the corresponding recombination path of them are displayed in Figure ~\ref{fig:Ga&O-FP_recomb}.  

Figure~\ref{fig:Ga&O-FP_recomb}(a) firstly shows four different vacancy sites ($\rm V1-V4$) around $\rm Ga_{iab1}$. $\rm V1-V3$ are all close to the interstitial at a distance around 4 \r A, while V4 is farther away at 6.36~\r A. 
By moving the interstitials to the vacancy sites, their recombination energy barriers $E_\mathrm{r}$ and the corresponding paths are presented in Figure~\ref{fig:Ga&O-FP_recomb}(c). The results of NEB simulations indicate that $E_\mathrm{r}$ of V1, V2, and V3 are approximately 0.4 eV. Such a low barrier suggests that these FPs are easier to recombine at low temperature. 
While for V4 the barrier increases to 0.83 eV with a higher reaction distance. 
{\JL The calculations show that the value of $E_\mathrm{r}$ is positively correlated to the reaction distance of $\rm Ga_{ia}$ and $\rm V_{Ga}$.}

For $\rm Ga_{ia}$-$\rm V_{Ga}$, the observation is quite different. Figure~\ref{fig:Ga&O-FP_recomb}(b) displays five different FPs and the recombination behavior of them are compared in Figure~\ref{fig:Ga&O-FP_recomb}(d). 
V1, V2, and V3 are vacancies located in the same Channel1 as the $\rm Ga_{ia}$, while V4 and V5 are not. 
{\HH By calculating $E_\mathrm{r}$ of each, all show higher $E_\mathrm{r}$ vaules, with the exception of V1. 
Despite V2 and V3 both being located in the same Channel1 of $\rm Ga_{ia}$, they present  significantly higher barriers ($>0.9$ eV) compared to V1.}
The overall values of $E_\mathrm{r}$ show that $\rm Ga_{ia}$-$\rm V_{Ga}$ are more difficult to recombine than $\rm Ga_{iab1}$-$\rm V_{Ga}$.  
We attribute this difference to the extra shared Ga1 site in $\rm Ga_{iab1}$-$\rm V_{Ga}$, which provides the interstitial atoms with more space to recombine. While in $\rm Ga_{ia}$-$\rm V_{Ga}$, the Ga interstitial is more difficult to move due to the denser atomic environment around it.

With respect to $\rm O_{sp\text{-}1site}$ FPs, nine O vacancies near $\rm O_{sp\text{-}1site}$ are selected to comprehensively compare with the nine Ga FPs, as shown in Figure~\ref{fig:Ga&O-FP_recomb}(e). V3, V4 and V6 show low $E_\mathrm{r}$ ($\sim0.2$ eV) which means that these vacancy sites have a strong attraction to the $\rm O_{sp\text{-}1site}$. These three vacancies are all located in the same channel (Channel1 and Channel3) for $\rm O_{sp\text{-}1site}$. For the vacancy sites that are not in the same channel, we also calculated their $E_\mathrm{r}$ values which are ranging from 0.35 to 0.82 eV. V5 exhibits the highest maximum $E_\mathrm{r}$ despite not having the furthest reaction distance. {\HH This trend contradicts the behavior observed in the Ga FPs. As illustrated in {\red the supplementary material Figure~S9}, the relationship between $E_\mathrm{r}$ and reaction distance for three types of FPs is different. The recombination behavior of O FPs demonstrates a comparatively lower $E_\mathrm{r}$ compared to Ga FPs. Even at the distance of 10 \r A, the relative lower $E_\mathrm{r}$ of O FPs indicate that oxygen of $\rm O_{sp\text{-}1site}$ breaks the bond easily and then occupies the vacancy sites.}

\section{Discussion} \label{sec:Discussion} 


{\HH TDEs in \ce{Ga2O3} were recently investigated using AIMD in Ref.~\cite{AIMD-2023TDEGa2O3}. The results showed that the minimal $E_\mathrm{d}$ values of Ga and O PKA are 28 eV and 14 eV, respectively. Compared with their findings, our large-scale MD simulations based our ML-IAPs indicate that the minimum $E_\mathrm{d}$ values of each PKA are very low, approximately ranging from 7 to 8 eV, as displayed in Figures~\ref{fig:TDE-plot} and \ref{fig:TDE-map}. However, both our MD and AIMD simulations consistently reveal that Ga PKAs have a higher mean $E_\mathrm{d}$ than O PKAs. Therefore, we compared the simulation parameters between two different methods in calculating $E_\mathrm{d}$ (as listed in {\red the supplementary material Table~S2}), including the system size, time step, and the number of simulated directions. Though AIMD simulations can calculate $E_\mathrm{d}$ efficiently without relying on empirical potentials, we still believe that MD simulations are the preferred method for obtaining statistically significant results when an accurate potential is available. Despite this, AIMD simulations remain valuable tools for investigating $E_\mathrm{d}$ along specific directions and they are able to explore the defect configurations while considering electronic effects.}

Previous AIMD~\cite{xiao2009threshold, xiInitioMolecularDynamics2018, AIMD-2023TDEGa2O3} and MD simulations~\cite{hePrimaryDamage102020, Jesper-byggmastar2019machine} both have shown that the value of $E_\mathrm{d}$ is highly dependent on the direction. Therefore, to accurately describe the mean TDE value, an adequate number of directions should be considered, especially for materials like $\beta$-\ce{Ga2O3} which has multiple nonequivalent atoms. Even if each PKA type is simulated over 1,000 directions to ensure statistical significance, it is still challenging to identify patterns in the entire TDE map of $\beta$-\ce{Ga2O3} in this work. Nordlund \textit{et al.}~\cite{nordlundLargeFractionCrystal2016} conducted a comprehensive study of the whole channeling map of several materials with cubic, FCC or diamond structure and they scanned angles at intervals of $1\degree$. A total of 300 million directions were carried out for materials with high symmetry. However, as displayed in Figure~\ref{fig:TDE-map}, the transformed symmetry of $\beta$-\ce{Ga2O3} still needs coverage of directions ranging from $\theta\in[0\degree,90\degree]$ and $\phi\in[0\degree,360\degree]$. It is speculated that more than $5\times90\times360$ directions are necessary to create a standard TDE map for $\beta$-\ce{Ga2O3}, which has a high computational cost.Additionally, it is important to exercise caution when considering directions in materials such as $\beta$-\ce{Ga2O3}, which possesses a monoclinic structure with a $\beta$ angle of $103.9\degree$. In this case, it should be noted that the direction of \hkl[100] is not perpendicular to the \hkl(100) plane, which can potentially lead to misleading interpretations. 

For the defects produced by PKAs, our MD simulations reveal that radiation damage leads to the generation of a great number of possible defects in this material and we have elucidated the underlying mechanisms through a dynamic process perspective.
Through extensive simulations involving a large system, we have identified new metastable configurations that have not been previously reported in DFT calculations. Specifically, our findings regarding Ga FPs indicate the presence of over ten such pairs, with a majority exhibiting split-interstitial configurations. This observation is partially consistent with recent Ga-related defect studies by Frodason \textit{et al.}~\cite{frodason2023migration}.
Their work reported that both Ga vacancies and interstitials exhibit split configurations, identifying three stable single interstitial sites (ib, ic, and if) and three split-interstitial sites of $\rm Ga_{iab1}$, $\rm Ga_{iad2}$ and $\rm Ga_{iac1}$.
Our observations align well with their findings of stable split-interstitial configurations.
However, the results of single site is not similar as only $\rm Ga_{ia}$-$\rm V_{Ga}$ is found during our TDE simulations.
We hypothesize that the presence of vacancies may influence the interstitial sites as seen in the similarity between $\rm Ga_{ia}$ and $\rm Ga_{if}$.
{\HH Furthermore, the recent DFT studies~\cite{frodason2023migration} have also shown that the $E_\mathrm{f}$ of split vacancies can be as low as that of single vacancies. However, limited to our developed detection methods, it is difficult to detect them now. Therefore, some of our identified Ga FPs may have different names for the  same configuration. For instance, $\rm Ga_{iab1}$-$\rm V_{Ga}$ can be also interpreted as $\rm Ga_{ia}$-$\rm V^{ib}_{Ga}$ when $\rm V_{Ga}$ of $\rm Ga_{iab1}$-$\rm V_{Ga}$ is close to $\rm Ga_{ib}$.}

The variety of Ga FPs underscore the complexity of Ga-related defects in this material, particularly under conditions of PKA impacts.
Notably, whether stable or metastable, a majority of these pairs consist of $\rm Ga_{ia}$, indicating a preference for $\rm Ga_{i}$ to localize within the largest channel.
Besides that, $\rm Ga_{ib}$ and $\rm Ga_{id}$ (in Channel2) exhibit higher formation probabilities than $\rm Ga_{ic}$ and $\rm Ga_{ie}$ (in Channel3), though they are both in two similar six-edge channels. From the data in the TDE maps of Ga sites in Figure~\ref{fig:TDE-map}(a) and (b), both Ga1 and Ga2 atoms show a preference for movement into Channel2 rather than Channel3. Therefore, the inconsistency in formation probabilities can be attributed to kinetic effects rather than thermodynamic effects. Furthermore, the transformation behaviors between these stable and metastable Ga FPs have low barriers, indicating that the stable defect system over longer time scales may be different and more complicated. 

{\HH For O FPs, the formation probability of $\rm O_{i}$ and $\rm O_{sp}$ FPs exhibit the different correlation with the specific type of oxygen sites.} With respect to $\rm O_{i}$ FPs, their formation probabilities are strongly influenced by the atomic sites of the vacancy rather than interstitial. Among the $\rm O_{i}$ FPs, $\rm O_{i}$-$\rm V_{O2}$ has the lowest production rate, approximately half that of the other two types.
This observation is not consistent with DFT simulations~\cite{Oxygen_ingebrigtsenImpactProtonIrradiation2019}, which have shown that $\rm V_{O2}$ has the lowest formation energy.
The structure of $\rm O_{i}$ and the TDE maps of the three O atoms both suggest that kinetic effects are the primary factor, as opposed to pure thermodynamic trends. Given that $\rm O_{i}$ is exclusively formed in the largest channel composed of O1 and O3 atoms, the movement of O2 into this Channel1 is more challenging, resulting in variations in the quantity of $\rm O_{i}$-$\rm V_{O2}$ FPs compared to the other types. For $\rm O_{sp}$ FPs, they display a contrasting trend as their formation is strongly influenced by the atomic sites of interstitial rather than vacancy.
Among all the O FPs, the $\rm O_{sp-1site}$ FP is the most commonly produced.
This preference is determined by the lowest formation energy, which is supported by our results and previous DFT simulations~\cite{Oxygen_ingebrigtsenImpactProtonIrradiation2019} indicating that $\rm O_{sp}$ tends to occupy the O1 site.
Furthermore, the investigation of the recombination barriers for three representative FPs reveals that O FPs exhibit lower recombination barriers, which is in good agreement with the findings from AIMD simulations~\cite{AIMD-2023TDEGa2O3}.
These results are also supported by previous experimental observations~\cite{azarovUniversalRadiationTolerant2023}, where O atoms remained stable after heavy ion irradiation while Ga atoms are displaced to attend in a phase transformation. 
Despite the fact that O atoms are more susceptible to displacement, their low recombination barrier indicates that the produced O-defects are not stable following radiation damage.

\section{Conclusions} \label{sec:Conclusions}

We conducted MD simulations employing ML-IAPs to investigate displacement events of low energy in $\beta$-\ce{Ga2O3}.
Utilizing our computationally fast tabGAP ML-IAP, more than five thousand random crystal directions for the five nonequivalent atoms in the $\beta$-\ce{Ga2O3} lattice have been simulated to determine statistically significant TDE values.
In terms of Ga atoms, Ga1 exhibits a higher mean TDE value (22.9 eV) compared to Ga2 (20.0 eV).
Analysis of their TDE maps reveals that the octahedral environment of Ga2 allows more space for displacement from the original sites.
Regarding oxygen atoms, the three O atoms exhibit similar distributions with mean TDE values ranging from 17.0 to 17.4 eV, however, their TDE maps demonstrate significant differences when considering the directions.
Subsequently, the final defective systems resulting from the TDE simulations are further analyzed using our developed defect identification methods.
More than ten types of Ga FPs are classified, with four of them identified as {\HH frequent} FPs accounting for 95\% of the total number of simulations.
The configurations of split interstitials in the largest channel are preferred for Ga interstitials.
Additionally, detailed calculations are performed on the O FPs composed of O interstitial and O-O split interstitial FP. $\rm O_{i}$-$\rm V_{O2}$ is the least common among $\rm O_{i}$ FPs due to difficulties in moving O2 into the largest channel while $\rm O_{sp\text{-}1site}$ is the most prevalent O FP owing to its low formation energy. 
NEB calculations for Ga and O FPs indicate that although O atoms are more easily displaced compared to Ga atoms, their lower recombination barrier allow a high probability of self-healing.
Our calculations are helpful to gain a deeper understanding of the {\JL radiation damage} and complex defect structures in $\beta$-\ce{Ga2O3}. 

\section*{Declaration of interests}

The authors declare that they have no financial or personal relationships with individuals or organizations that could potentially influence the work reported in this paper. 

\section*{Acknowledgements}

The authors would like to extend our sincere gratitude to the China Scholarship Council (CSC) for funding the studies of H. He in University of Helsinki.
J. Zhao acknowledge the National Natural Science Foundation of China under Grant 62304097; Guangdong Basic and Applied Basic Research Foundation under Grant 2023A1515012048; Shenzhen Fundamental Research Program under Grant JCYJ20230807093609019. J. Byggmästar acknowledges financial support from the Research Council of Finland flagship programme: Finnish Center for Artificial Intelligence (FCAI). The computational resources are provided by the Finnish Computing Competence Infrastructure (FCCI); IT Center for Science, CSC, Finland; and the Center for Computational Science and Engineering at the Southern University of Science and Technology. The authors are also grateful for the valuable discussions by Mr. J. Wu, Mr. J. Zhang, Dr. H. Liu, and Dr. I. Makkonen at the University of Helsinki. 

\section*{Supplementary Material}
Supplementary Material is available for this paper at \url{https://doi.org/...}



\bibliography{main}









\end{document}